\documentclass[%
preprint,
superscriptaddress,
shortbibliography,
 amsmath,amssymb,
 aps,
]{revtex4-1}

\usepackage{graphicx}
\usepackage{dcolumn}
\usepackage{bm}
\usepackage{tabularx}
\usepackage{hyperref}
\begin{document}

\title{Supercurrent in ferromagnetic Josephson junctions with heavy metal interlayers}

\author{Nathan~Satchell}
\affiliation{Department of Physics and Astronomy, Michigan State University, East Lansing, MI 48912, USA}
\affiliation{School of Physics and Astronomy, University of Leeds, Leeds, LS2 9JT, United Kingdom}

\author{Norman~O.~Birge}
\email{birge@pa.msu.edu}
\affiliation{Department of Physics and Astronomy, Michigan State University, East Lansing, MI 48912, USA}

\date{\today}

\begin{abstract}

The lengthscale over which supercurrent from conventional BCS, $s$-wave, superconductors (\textit{S}) can penetrate an adjacent ferromagnetic (\textit{F}) layer depends on the ability to convert singlet Cooper pairs into triplet Cooper pairs. Spin aligned triplet Cooper pairs are not dephased by the ferromagnetic exchange interaction, and can thus penetrate an \textit{F} layer over much longer distances than singlet Cooper pairs. These triplet Cooper pairs carry a dissipationless spin current and are the fundamental building block for the fledgling field of superspintronics. Singlet-triplet conversion by inhomogeneous magnetism is well established. Here, we describe an attempt to use spin-orbit coupling as a new mechanism to mediate singlet-triplet conversion in \textit{S--F--S} Josephson junctions. We report that the addition of thin Pt spin-orbit coupling layers in our Josephson junctions significantly increases supercurrent transmission, however the decay length of the supercurrent is not found to increase. We attribute the increased supercurrent transmission to Pt acting as a buffer layer to improve the growth of the Co \textit{F} layer.


\end{abstract}


\maketitle

\section{Introduction}

In nature there are very few examples of materials exhibiting simultaneously superconducting (\textit{S}) and ferromagnetic (\textit{F}) properties, due to the competition between the order parameters. Breakthroughs in materials engineering and nanolithography techniques in the last two decades have enabled the synthesis of artificial heterostructures, where two or more layers are in direct electronic contact, revealing a wealth of new physics at \textit{S--F} interfaces \cite{eschrig_spin-polarized_2011, linder_superconducting_2015, 0034-4885-78-10-104501}. Exploitation of this new physics has led to advances in the emerging field of superspintronics, which offers a new class of highly energy efficient devices, most promisingly cryogenic memory elements based on ferromagnetic Josephson junctions \cite{bell_controllable_2004, goldobin_memory_2013, 6377273, baek_hybrid_2014, niedzielski_use_2014, gingrich_controllable_2016, PhysRevB.96.224515,niedzielski2017spin, dayton2017experimental,8359341}.

In the simplest case of a normal metal (\textit{N}) in \textit{S--N--S} Josephson junctions the critical current of the junction ($I_c$) will decay slowly with increasing \textit{N} thickness, on a typical lengthscale $\xi_N \approx $100's~nm \cite{PhysRevB.63.064502,RevModPhys.77.935}. When the \textit{N} layer is replaced by a ferromagnet, \textit{S--F--S}, singlet Cooper pairs entering the \textit{F} layer are dephased and gain an oscillatory term with \textit{F} layer thickness, driven by the same physics as the predicted Fulde-Ferrell-Larkin-Ovchinnikov (FFLO) state \cite{PhysRev.135.A550, larkin_nonuniform_1965, buzdin1982critical, PhysRevB.55.15174, RevModPhys.77.935}. This oscillation results in a series of transitions from a zero to $\pi$ ground-state phase difference across the Josephson junction. Experimentally, oscillations in $I_c$ are observed with increasing \textit{F} layer thickness \cite{PhysRevLett.89.137007, PhysRevLett.96.197003}, typically over a lengthscale $\xi_F \approx $1-5~nm for a strong ferromagnet \cite{PhysRevLett.89.187004, PhysRevLett.97.177003}. With the addition of spin mixing layers on either side of the \textit{F} layer it is possible to create the so-called long ranged triplet component (LRTC) \cite{PhysRevLett.86.4096, RevModPhys.77.1321}. Unlike singlet Cooper pairs, the LRTC is not dephased by the ferromagnet and can therefore penetrate further into the \textit{F} layer than the singlet component, typically $\xi_{\text{LRTC}} \approx $10's-100's~nm \cite{keizer2006spin, PhysRevLett.96.157002, robinson2010controlled, PhysRevB.82.100501, PhysRevLett.104.137002}. Experimentally, the LRTC is generated reliably through the addition of \textit{F', F''} ferromagnetic layers in \textit{S--F'--F--F''--S} Josephson junctions, where \textit{F', F''} have either intrinsic magnetic inhomogenity (for example Ho \cite{PhysRevLett.96.157002, robinson2010controlled}) or magnetic inhomogenity which is engineered in multilayered structures \cite{PhysRevLett.104.137002, doi:10.1063/1.3681138, PhysRevLett.116.077001}. It is known theoretically, however, that this spin mixing layer need not be a ferromagnet and there are several proposals indicating that spin-orbit coupling can act as a source for the LRTC \cite{doi:10.1063/1.4743001, PhysRevLett.110.117003, PhysRevB.89.134517, Konschelle2014, PhysRevB.92.024510, alidoust2015spontaneous, alidoust2015long, jacobsen2016controlling}.

Two recent experimental reports study the \textit{S}--\textit{F} proximity effect with the addition of spin-orbit coupling layers. Banerjee \textit{et al.}, measure $T_c(H)$ characteristics of Nb/Pt($d_{\text{Pt}}$)/Co/Pt multilayers with $0 \leq d_{\text{Pt}} \leq 2$~nm \cite{banerjee2017controlling}. The changes in $T_c(H)$ are attributed to spin-orbit coupling mediating singlet to triplet conversion. The results are presented in analogy to the triplet spin valve, where the presence of triplet correlations modifies the proximity effect, introducing measurable signatures in $T_c(H)$ \cite{fominov2010superconducting}. Jeon \textit{et al.} study ferromagnetic resonance (FMR) in $N_{\text{SOC}}$/Nb/Py/Nb/$N_{\text{SOC}}$ films, where the $N_{\text{SOC}}$ layers (normal metals with strong spin-orbit coupling) are Ta, W and Pt \cite{jeon2018enhanced}. With the addition of $N_{\text{SOC}}$ layers an enhancement in the spin pumping into the Nb \textit{S} layers is observed. This is the opposite behavior to what is expected for a superconductor, and is attributed to the LRTC. It is clear that the presence of spin-orbit coupling plays an important role in both experiments.

In this work we describe a different approach to generating the LRTC using a normal metal with strong spin-orbit coupling ($N_{\text{SOC}}$) in Josephson junctions. We compare the transport properties of two sets of Josephson junctions; \textit{S--F--S} and \textit{S}--$N_{\text{SOC}}$--\textit{F}--$N_{\text{SOC}}$--\textit{S}, where \textit{S} is Nb, \textit{F} is a Co/Ru/Co synthetic antiferromagnet (SAF), and $N_{\text{SOC}}$ is Pt (which has been shown in previous works to have strong Rashba spin-orbit coupling with Co due to broken inversion symmetry \cite{miron2011perpendicular,haazen2013domain,PhysRevB.90.020402}). It is our proposal that only the set of samples containing two $N_{\text{SOC}}$ layers should display properties consistent with the generation of the LRTC. In $N_{\text{SOC}}$-containing samples the decay of supercurrent transmitted through the \textit{F} layer should take place over a longer length scale than in the non-$N_{\text{SOC}}$ samples, which only contain the short ranged supercurrent components (namely singlet and $m_s=0$ triplet Cooper pairs).


\section{Methods}
The films are deposited using DC sputtering in a vacuum system with base pressure of $2 \times 10^{-8}$~Torr and partial water pressure of $3 \times 10^{-9}$~Torr after liquid nitrogen cooling. The samples are grown on 0.5~mm thick Si substrates which have a native oxide layer. A uniform 200~Oe magnetic growth field is applied to the substrates. Growth is performed at an approximate Ar pressure of 2~mTorr, at a typical growth rate of 0.4~nm~s$^{-1}$ for Nb and 0.1-0.2~nm~s$^{-1}$ for the other materials. Growth rates are calibrated using an \textit{in situ} crystal film thickness monitor and checked by fitting to Keissig fringes obtained by low angle X-ray reflectometry on reference samples. All layer thicknesses (in brackets) are in nm. The bottom superconducting electrode is a multilayer [Nb(25)/Al(2.4)]$_3$/Nb(20) which grows considerably smoother than single layer Nb of comparable total thickness \cite{PhysRevB.85.214522, doi:10.1063/1.368037, doi:10.1063/1.363459}. The bottom electrode, ferromagnetic layers, any normal metal interlayers, and a capping bilayer Nb(5)/Au(15) are grown without breaking vacuum.

For electrical transport measurements, films are patterned into circular Josephson junctions of diameter; 12, 24 and 48~$\mu$m using standard photolithography and ion milling methods, described in previous work \cite{PhysRevB.85.214522}. Once the Josephson junctions are defined, the samples are returned to the DC sputtering system and the top Au(15) layer is ion milled \textit{in situ} thus recovering a very clean interface before depositing the top superconducting electrode, Nb(150). 

Magnetization loops of sister sheet film samples are measured using a Quantum Design SQUID VSM magnetometer at 10~K. X-ray diffraction (XRD) of sheet film samples are measured using a Bruker D8 diffractometer at room temperature. Electrical transport is performed using a conventional four-point-probe measurement configuration at 4.2~K, employing the low noise electrical transport system described in reference \cite{PhysRevB.96.224515}. Our system can resolve 6~pV, which is taken as a resolution limit where appropriate. For all transport measurements the field is applied parallel to the sample's plane, and samples are measured in the as-grown magnetic state (which is set by the growth field). 

\section{Magnetic Characterization}

\begin{figure} 
\includegraphics[width=1\columnwidth]{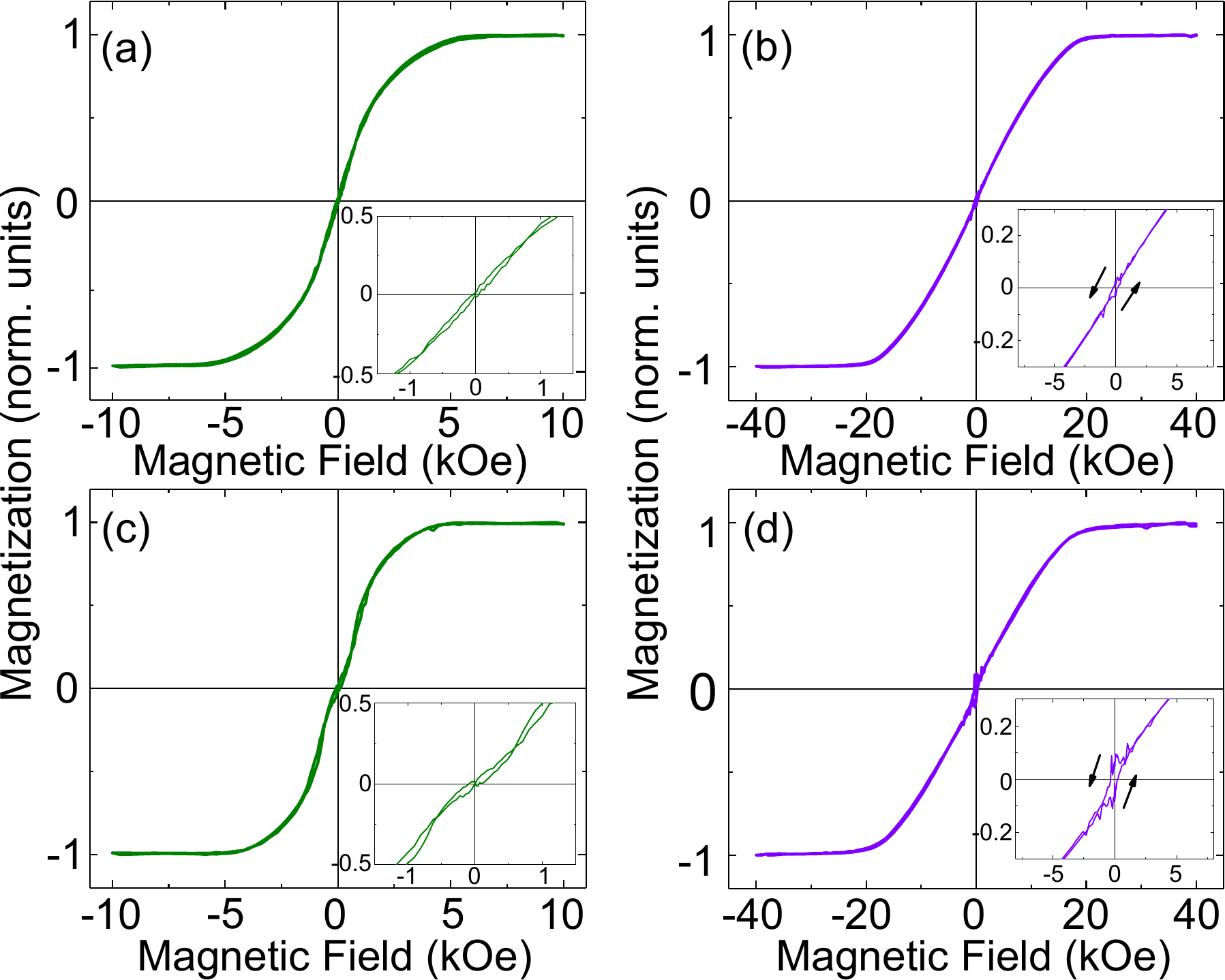} 
\caption{Magnetic hysteresis loops acquired at a temperature of 10~K. (a,b) for \textit{S--F--S} type samples with the applied field oriented (a) in and (b) out of the sample plane and (c,d) for \textit{S}--Pt(0.5)--\textit{F}--Pt(0.5)--\textit{S} type samples (c) in and (d) out of the sample plane. \textit{F} is a Co(5)/Ru(0.6)/Co(5) multilayer in all cases.   The diamagnetic contribution from the substrate has been subtracted and data are normalized by the saturated value of magnetization. Insets show the low field switching. \label{loops}}
\end{figure}


The magnetization versus field data are shown in FIG.~\ref{loops}, for (a,b) \textit{S--F--S} and (c,d) \textit{S}--$N_{\text{SOC}}$--\textit{F}--$N_{\text{SOC}}$--\textit{S} samples at 10~K, where \textit{F} is a Co(5)/Ru(0.6)/Co(5) multilayer in all cases and $N_{\text{SOC}}$ is Pt(0.5). The choice of Pt thickness here is dictated by the transport measurements to follow. Both samples behave as synthetic antiferromagnets (SAFs). The application of an applied field in-plane (a,c) causes a spin-flop transition and then rotates the magnetizations into the direction of the applied field, saturating at about 5~kOe. Removing the field causes the Co layers in the SAF to relax antiparallel w.r.t. each other and perpendicular w.r.t. to the applied field, hence zero remanent magnetization is observed. The spin-flop transition in similar samples was confirmed in previous work by polarized neutron reflectometry \cite{PhysRevLett.108.127002}. In FIG.~\ref{loops} (c), the addition of the Pt(0.5) interlayers has caused a slight reduction in the low-field susceptibility, between about -1 and 1~kOe (highlighted in the figure inset). This implies that there are magnetic phases present with different coercivities, which we can understand if the surface moments couple with the Pt layer, modifying the local anisotropy of the Co/Pt interface. Or alternatively, this may be a direct signature of the spin-flop transition in this sample.

The addition of Pt at the Co interface may induce an additional out-of-plane magnetic anisotropy \cite{PhysRevLett.81.5229}. The response of both samples to an out-of-plane applied field, as shown in FIG.~\ref{loops} (b,d), indicates that the magnetic anisotropy of our samples lie predominantly in-plane as very little out-of-plane remanent magnetization is observed. The sample containing Pt interlayers may have a small out-of-plane component, shown in FIG.~\ref{loops} (d) and inset. This is not surprising given the thickness of Co and Pt layers in this study compared to the previous work of Shepley \textit{et al.} \cite{shepley2015modification}; where an out-of-plane magnetic anisotropy is achieved for Pt thickness of 2.5~nm and corresponding Co thicknesses in the range 0.85-1.0~nm. The reorientation transition from predominant out-of-plane to in-plane magnetic anisotropy is found at Co thickness 1.1~nm \cite{shepley2015modification}. 

We do not expect the slight differences in magnetic switching between samples with and without Pt interlayers (FIG.~\ref{loops}) to affect our transport measurements, which are performed in the as-grown magnetic state.


\section{Electrical Transport}

Typical \textit{I}--\textit{V} curves and Fraunhofer $I_c$ ($B$) patterns for each Josephson junction size along with the collated $AR_N$ (area times the normal state resistance) are shown in the Supplemental Materials \cite{NoteX}.


\subsection{$S$--Pt($d_{\text{Pt}}$)--Co(5)/Ru(0.6)/Co(5)--Pt($d_{\text{Pt}}$)--$S$}


\begin{figure} 
\includegraphics[width=0.95\columnwidth]{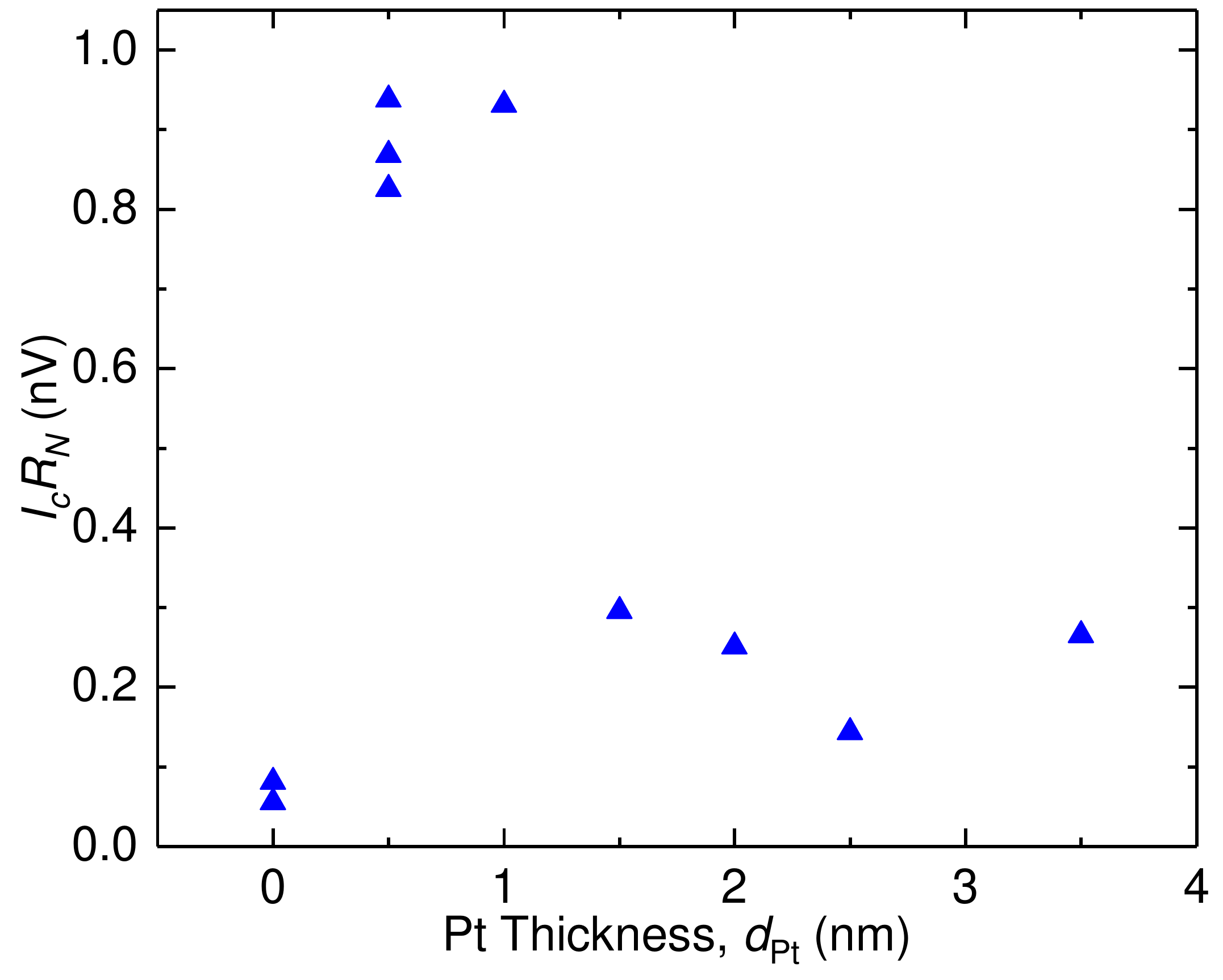} 
\caption{Product of critical current times normal state resistance vs Pt interlayer thickness ($d_{\text{Pt}}$) for Josephson junctions of the form $S$--Pt($d_{\text{Pt}}$)--Co(5)/Ru(0.6)/Co(5)--Pt($d_{\text{Pt}}$)--$S$. Each data point represents one Josephson junction and the uncertainty in determining $I_cR_N$ is smaller than the data points. \label{Ptx}}
\end{figure}

We first consider the transport characteristics of the set of samples; $S$--Pt($d_{\text{Pt}}$)--Co(5)/ Ru(0.6)/Co(5)--Pt($d_{\text{Pt}}$)--$S$, where $d_{\text{Pt}}=0-3.5$~nm. A total Co thickness of 10~nm is chosen as it is known from previous works to be a thickness where LRTC and non-LRTC samples show obvious difference in $I_c$ (approximately an order of magnitude) \cite{PhysRevLett.104.137002}. FIG.~\ref{Ptx} shows the result of this study, where characteristic junction voltage $I_cR_N$ (the product of the maximum measured $I_c$ in the Fraunhofer pattern times normal state resistance) is plotted for a series of samples with increasing $d_{\text{Pt}}$. Without the Pt, our Josephson junctions have an expected small $I_cR_N$. With the addition of $d_{\text{Pt}}=0.5$~nm (less than two monolayers) we see a large enhancement in $I_c$ and no change in $R_N$, increasing the characteristic junction voltage by approximately an order of magnitude. This high $I_cR_N$ remains approximately constant for $d_{\text{Pt}}=1.0$~nm Pt interlayers, but for increasing Pt thicknesses beyond this there is a sharp drop in $I_cR_N$, where the critical current of our Josephson junctions appears to fall back towards the value without any Pt.

\subsection{$S$--Pt(0.5)--Co($d_{\text{Co}} /2$)/Ru(0.6)/Co($d_{\text{Co}} /2$)--Pt(0.5)--$S$}

\begin{figure} 
\includegraphics[width=1\columnwidth]{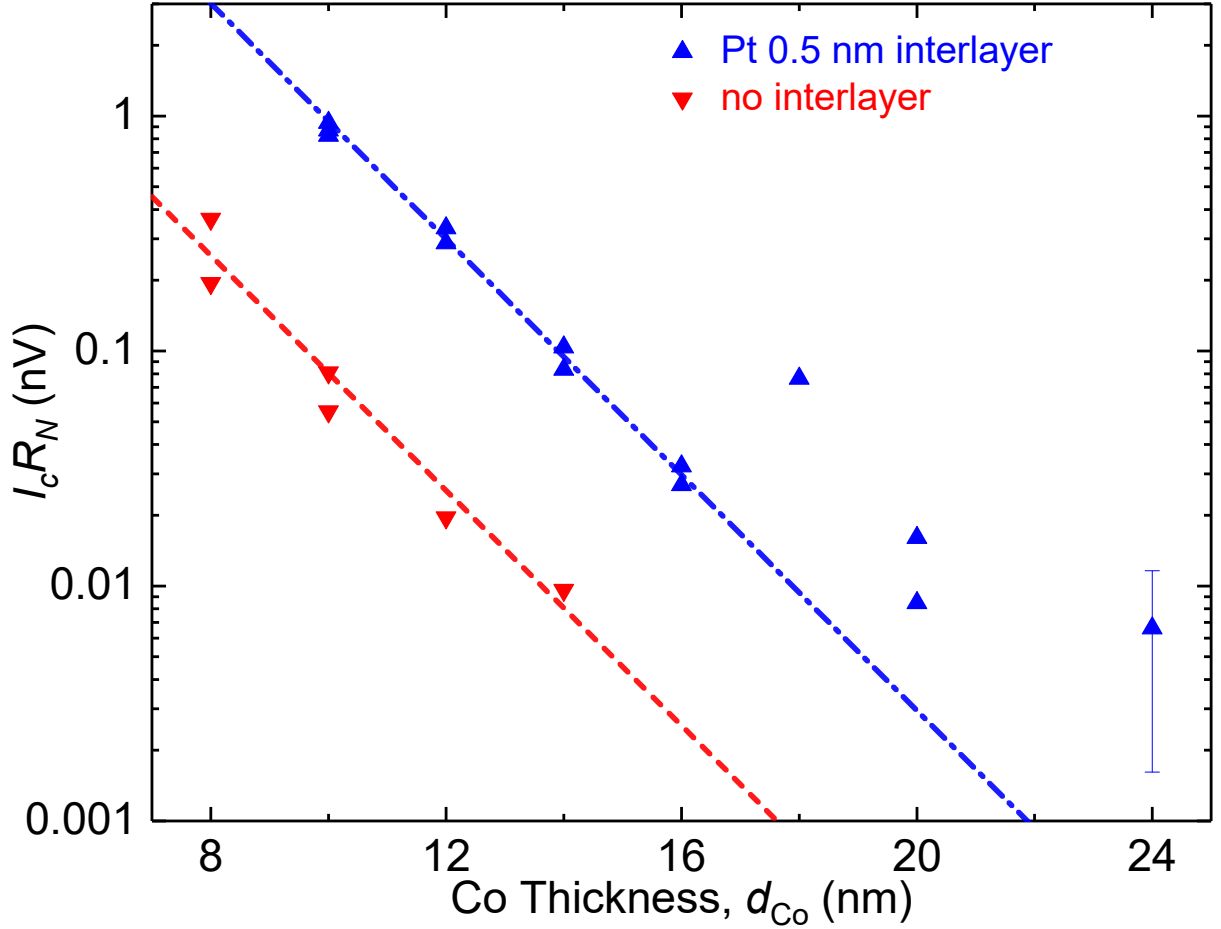} 
\caption{Product of critical current times normal state resistance vs total Co thickness ($d_{\text{Co}}$) for Josephson junctions of the form; blue triangles $S$--Pt(0.5)--Co($d_{\text{Co}} /2$)/Ru(0.6)/Co($d_{\text{Co}} /2$)--Pt(0.5)--$S$ and red inverted triangles $S$--Co($d_{\text{Co}} /2$)/Ru(0.6)/Co($d_{\text{Co}} /2$)--$S$. Each data point represents one Josephson junction and where no error bars are shown, the uncertainty in determining $I_cR_N$ is smaller than the data point. The lines are fits to simple exponential decay in the range 8~nm $\leq d_{\text{Co}} \leq $ 16~nm, with decay lengths of $1.73 \pm 0.07$~nm and $1.7 \pm 0.2$~nm, respectively.\label{Cox}}
\end{figure}

With a fixed Pt thickness of 0.5~nm, guided by FIG.~\ref{Ptx}, we next vary the thickness of the \textit{F} layer in $S$--Pt(0.5)--Co($d_{\text{Co}} /2$)/Ru(0.6)/Co($d_{\text{Co}} /2$)--Pt(0.5)--$S$ and $S$--Co($d_{\text{Co}} /2$)/Ru(0.6)/ Co($d_{\text{Co}} /2$)--$S$ samples to compare the decay length of the supercurrent with and without the Pt interlayer. For these samples $I_cR_N$ as a function of total Co thickness $d_{\text{Co}}$ is plotted in FIG.~\ref{Cox}. Considering first the data for samples without Pt, where we expect to only have short ranged supercurrent components inside the $F$ layer, we observe the expected rapid decay in $I_cR_N$ \cite{PhysRevB.80.020506}. Fitting to a simple exponential decay the decay length is found to be $1.7 \pm 0.2$~nm. This decay length is longer than found in our previous study of a Co SAF, where the comparable samples have decay length $1.18\pm0.05$~nm \cite{PhysRevB.80.020506}. The differences between this work and the previous work are discussed later. Samples thicker than $d_{\text{Co}}=14$~nm were fabricated but show no measurable critical current.

The samples containing Pt(0.5) interlayers can be described in two regimes. In the Co thickness range 10~nm $\leq d_{\text{Co}} \leq $ 16~nm the supercurrent is found to decay exponentially with decay length $1.73 \pm 0.07$~nm. This is identical within our experimental uncertainty to the decay length of samples without Pt interlayers and suggests that short ranged supercurrent components dominate transport in our Josephson junctions. The $I_cR_N$ product for the Pt interlayer samples is, however, consistently about one order of magnitude higher than the samples without the Pt interlayer. This suggests that the Pt has a role in the transmission of these short ranged supercurrent components. 

Although the $AR_N$ product of the $d_{\text{Co}}=18$~nm Josephson junction appears consistent with the other samples measured in this work, we believe this data point to be an anomaly. The other five Josephson junctions on this substrate showed electrical shorts when measured, indicating a failure in the lithographic processing. Samples showing electrical shorts were rejected from this study. A junction with an electrical short does not display the characteristic Fraunhofer $I_c$(B) pattern, rather the $I_c$ is found to decay monotonically with B (it is not a Josephson junction). While it is included in FIG.~\ref{Cox} for completeness, the $d_{\text{Co}}=18$~nm sample is not used in the analysis of decay lengths. A second reason for rejecting samples in this study is when the $AR_N$ product is considerably outside of the spread of $AR_N$ presented in Fig 1 of the Supplemental Materials \cite{NoteX}. This only happened on one set of samples which had increased $R_N$, where the microscope later revealed residual resist in the area of the junctions. A $d_{\text{Co}}=8$~nm sample containing Pt(0.5) interlayers did not survive fabrication processing.

In the second regime, upon increasing the Co thickness further it is found that unlike the samples without Pt, a small residual supercurrent is observable in Josephson junctions with Co thickness $d_{\text{Co}}=20$~nm and $d_{\text{Co}}=24$~nm. The $d_{\text{Co}}=20$~nm sample produces a clear critical current and Fraunhofer pattern, giving us confidence that the $I_cR_N$ of this sample is well above what is expected from the decay of short ranged supercurrent components, depicted by the blue line. The $d_{\text{Co}}=24$~nm sample shows some evidence for non-zero critical current, however the Fraunhofer pattern is not well defined, and the $I_cR_N$ product is at the resolution of our instrument (approximately 6~pV), leading to a large uncertainty in this value. We have included data from the $d_{\text{Co}}=20$ and 24~nm samples in FIG 4 of the Supplemental Materials \cite{NoteX}. The decay length in this regime (where short ranged supercurrent components are vanishingly small) appears much longer, but we do not have enough data points to place a meaningful quantitative value on the decay.

\subsection{Control Samples}


\begin{figure} 
\includegraphics[width=1\columnwidth]{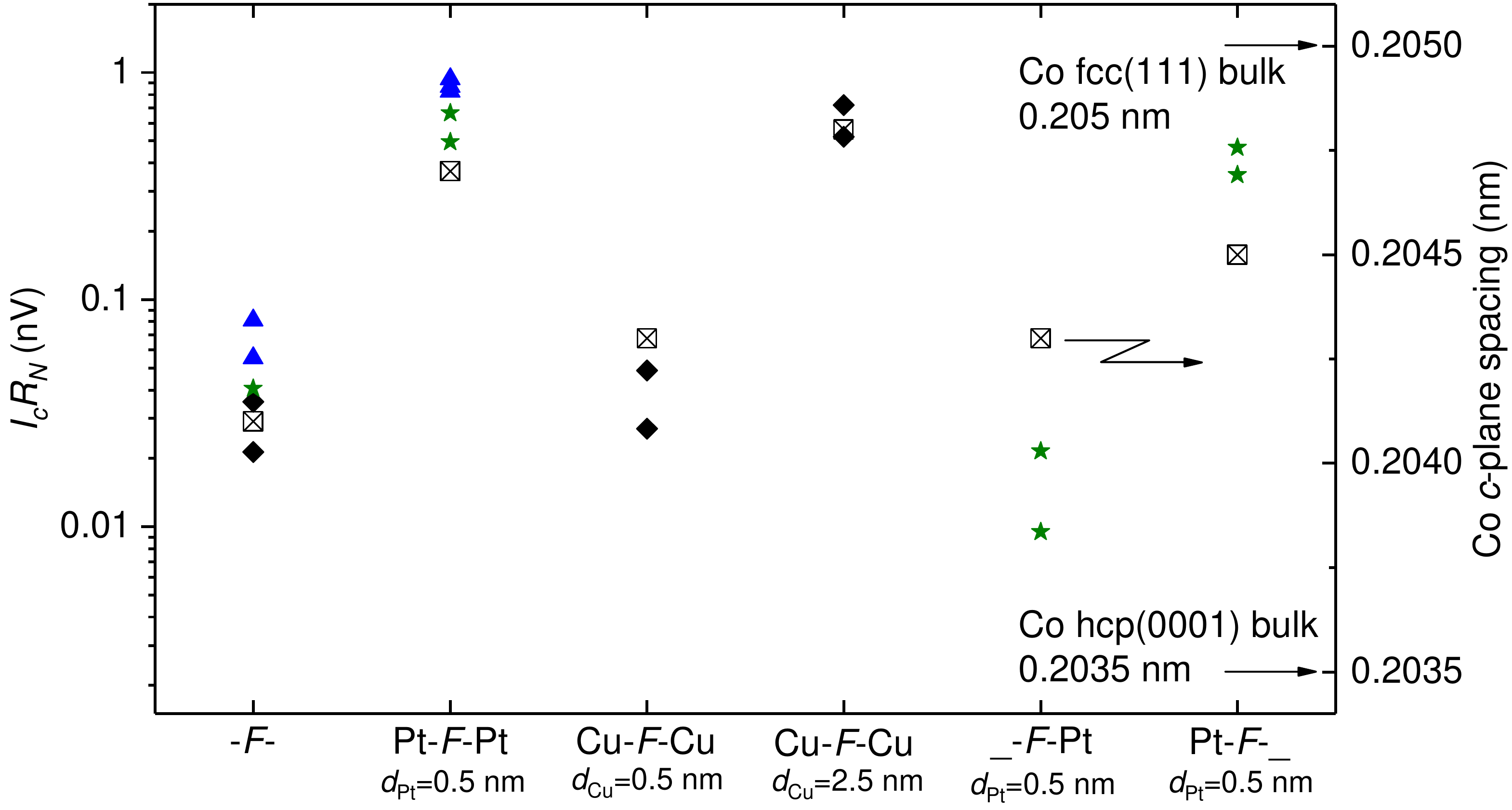} 
\caption{Product of critical current times normal state resistance (solid symbols) and Co $c$-plane spacing (open symbols) vs selected sample for; symmetric Josephson junctions containing no interlayers, two Pt(0.5) interlayers or two Cu(0.5, 2.5) interlayers, and asymmetric Josephson junctions which contain only one Pt(0.5) interlayer. Each symbol corresponds to a set of samples grown in the same vacuum cycle, and our run-to-run variation in $I_cR_N$ is visible in the -\textit{F}- only samples. All Josephson junctions contain a Co(5)/Ru(0.6)/Co(5) \textit{F} layer. Each solid data point represents one Josephson junction and the uncertainty in determining $I_cR_N$ is smaller than the data points. The open symbols show the Co $c$-plane spacing obtained by x-ray diffraction measurements on sheet films described in the text. \label{Cux}}
\end{figure}

As a first control measurement we consider the properties of a set of samples where we replace the Pt interlayer with a normal metal $N$, Cu. Cu is not expected to contribute much to spin-orbit coupling in our system and the supercurrent carrying properties of Cu are well characterized. It is known that supercurrent decay through Cu is very slow due to to the long electron mean-free-path \cite{PhysRevB.63.064502}. Additionally, Cu is already widely implemented as a normal metal interlayer in \textit{S--F} systems, where it is added into multilayer stacks as a buffer layer to improve growth conditions, and to decouple multiple \textit{F} layers where independent switching of each \textit{F} layer is desirable \cite{PhysRevB.80.020506, PhysRevB.89.184502, PhysRevB.91.214508}. 

FIG.~\ref{Cux} includes $I_cR_N$ for the set of samples; $S$--Cu($d_{\text{Cu}}$)--Co(5)/Ru(0.6)/Co(5)--Cu($d_{\text{Cu}}$)--$S$, where $d_{\text{Cu}}=0$, 0.5 and 2.5~nm (black diamonds). With no interlayer, once again a small $I_cR_N$ is observed. The slightly lower $I_cR_N$ of this sample set (compared to the previous sample set, blue triangles) is mostly likely run-to-run variation between samples grown in different vacuum cycles. Adding 0.5~nm of Cu does not significantly improve the transmission of supercurrent through the \textit{F} layer, unlike Pt. With an increase in Cu thickness to 2.5~nm, an increase in $I_cR_N$ to approximately the value of the Pt(0.5) interlayer samples is observed. This increase with thick Cu interlayers is expected from previous work \cite{PhysRevB.80.020506} and is due to improved structural properties of the Co, a mechanism outlined further in Section~\ref{discussion}.

The next control samples are asymmetric Josephson junctions containing only one Pt(0.5) interlayer, olive green stars on FIG.~\ref{Cux}. We find that the interface on which the Pt is grown is of key importance to the observed increase in $I_cR_N$. When the Pt layer is grown only on the top interface ($S$--Co(5)/Ru(0.6)/Co(5)--Pt(0.5)--$S$) a small \emph{reduction} in $I_cR_N$ is observed. On the other hand, when the Pt layer is grown only on the bottom interface ($S$--Pt(0.5)--Co(5)/Ru(0.6)/Co(5)--$S$) a large \emph{increase} in $I_cR_N$ is observed, almost recovering the value for the symmetric Pt(0.5) samples (replotted blue triangles on FIG.~\ref{Cux}).

We also perform x-ray diffraction measurements on sheet films for each of the selected samples plotted in FIG.~\ref{Cux}. We simplify the structure of the sheet films so the seed and capping layer is Nb(5) and remove the Ru layer to make the data interpretation easier. The main structural peak for the Co(10) \textit{F} layer is easily visible in the x-ray diffraction pattern and used to determine the Co $c$-plane spacing for each sample. We find that the interlayers used in this work subtly change the growth morphology of the Co layer. The Pt(0.5) and Cu(2.5) interlayers push the Co towards the bulk fcc(111) phase, which corresponds to the Josephson junctions with highest $I_cR_N$. With no interlayers or with Cu(0.5) interlayers the Co appears to contain a mixture of fcc(111) and hcp(0001) phases, which corresponds to the Josephson junctions with lower $I_cR_N$. The asymmetric samples follow the same trend.



\section{Discussion}\label{discussion}




Direct evidence for the presence of a LRTC of superconductivity in a Josephson junction is the slower decay of $I_cR_N$ with increasing \textit{F} layer thickness. This ``smoking gun'' is reliably observed in samples of the form \textit{S--F'--F--F''--S} where \textit{F', F''} are aligned perpendicular to \textit{F} \cite{PhysRevLett.104.137002, doi:10.1063/1.3681138, PhysRevLett.116.077001}. This increase in supercurrent decay length is often (but not necessarily) accompanied by an increase in $I_cR_N$ for LRTC samples compared to non-LRTC samples for typical \textit{F} layer thicknesses in experiments. In this work for the Co thickness range 8~nm $\leq d_{\text{Co}} \leq $ 16~nm we observe increase in $I_cR_N$, without the corresponding increase in decay length. We therefore do not believe the increased transmission of supercurrent into the Co layers for the thickness range 8~nm $\leq d_{\text{Co}} \leq $ 16~nm can be attributed to the presence of a LRTC generated by spin-orbit coupling.

For our Josephson junctions with Co thickness $d_{\text{Co}}=20$~nm and $d_{\text{Co}}=24$~nm we measure a very small residual supercurrent. The measured $I_cR_N$ of these junctions is well above what is expected from the decay of short ranged supercurrent components, the blue line in FIG.~\ref{Cox}. The decay length of superconductivity for these Co thicknesses appears to be much longer than for thinner Co. These observations offer the best evidence in this work that spin-orbit coupling can mediate the conversion of Cooper pairs to the LRTC. This residual supercurrent is many orders of magnitude lower than comparable samples with \textit{F', F''} LRTC generating interlayers \cite{PhysRevLett.104.137002}. This suggests at a minimum that \textit{if} we attribute these observations to spin-orbit coupling generating a LRTC that the conversion efficiency of such a mechanism is poor. We must caution that the supercurrent observed in these junctions is only just above the resolution limit of our measurement apparatus.


We next address the values of $I_cR_N$ and supercurrent decay length obtained in this work compared to the previous work of Khasawneh \textit{et al.}, which also considers the transmission of singlet supercurrent through a Co/Ru/Co SAF \cite{PhysRevB.80.020506}. The reported $I_cR_N$ for samples containing no interlayers is approximately an order of magnitude lower in this work compared to the previous work. For Josephson junctions in this work containing either Cu(2.5) or Pt(0.5) interlayers the $I_cR_N$ is also about an order of magnitude lower, compared to the Cu(5) interlayers of Khasawneh \textit{et al.}. The decay length inside the Co for the samples with no interlayers also differs, but in the opposite way to $I_cR_N$. In this work the decay length is improved to $1.7 \pm 0.2$~nm compared to the previous study, which reports a decay length of $1.18\pm0.05$~nm. There are three main differences between the Josephson junctions in this work and those studied by Khasawneh \textit{et al.} which could contribute to the lower $I_cR_N$ and longer decay length. Firstly, this work uses the smoother [Nb(25)/Al(2.4)]$_3$/Nb(20) multilayer and not thick single layer Nb(150) as a bottom electrode. One difference of note is that the multilayer has a reduced $T_c$ (8.0~K compared to 9.0~K). Secondly, the capping layers on the initial stack in this work are Nb(5)/Au(15) compared to Nb(25)/Au(15) in the previous work. Thirdly, when we introduce Cu here we only study 0.5 and 2.5~nm thick Cu interlayers, in the previous work 5~nm thick Cu is used. A more detailed comparison is made in the Supplemental Materials \cite{NoteX}. In our future work we will study these differences systematically, as this might have technological implications for superspintronic devices. 


In the previous work of Khasawneh \textit{et al.} it is found that the addition of Cu(5) on either side of the Co/Ru/Co SAF increases the decay length of supercurrent from $1.18\pm0.05$~nm to $2.34\pm0.08$~nm, accompanied with an increase in $I_cR_N$ \cite{PhysRevB.80.020506}. In that work the improvement was attributed to Cu changing the growth characteristic of Co, improving the mean-free-path. Here, we are able to offer some additional insight into this result. We grow two thicknesses of Cu, 0.5 and 2.5 nm, and measure both the $I_cR_N$ of the Josephson junction and the Co $c$-plane spacing in corresponding sheet films. It is found that Cu(0.5) interlayers are not thick enough to affect the growth characteristic of Co compared to the samples with no interlayers, and correspondingly no improvement in the transmission of supercurrent is observed (FIG.~\ref{Cux}). The Co $c$-plane spacings for both the Cu(0.5) sample and the no interlayer sample fall between the bulk fcc(111) and hcp(0001) values, which we interpret as the Co initially growing in an fcc phase, before relaxing to hcp. This is a known phenomena for Co and creates stacking faults and dislocations in the Co grains \cite{PhysRevLett.115.056601}. In previous high resolution transmission electron microscopy studies it is found that Cu will grow in a nonequilibrium bcc phase on bcc Nb for Cu thicknesses up to 1.2~nm \cite{JACE:JACE1673,doi:10.1063/1.119611,doi:10.1063/1.371342}. In this bcc phase the lattice parameter of Cu is $0.328 \pm 0.007$~nm (close to the bulk bcc Nb lattice parameter of 0.331~nm). This explains why the Cu(0.5) and no interlayer samples behave so similarity. It is possible that the presence of stacking faults and dislocations are responsible for the lower critical current in these samples.

Upon growing Cu(2.5) interlayers, we recover the higher $I_cR_N$ values also found for the Pt(0.5) interlayers samples. By 2.5~nm the Cu has recovered fcc growth, and the positive influence this has on the Co growth is observed in both the improved $I_cR_N$ and $c$-plane spacing. From our XRD measurements it is clear that both Pt(0.5) and Cu(2.5) interlayers have a similar effect on the growth characteristic of the Co layer (FIG.~\ref{Cux} open symbols). In these samples the Co $c$-plane spacing recovers almost the bulk fcc(111) value. This suggests two things; firstly, unlike Cu, Pt does not form a nonequilibrium bcc phase for very thin layers grown on bcc Nb \cite{doi:10.1063/1.2734378, doi:10.1063/1.3553414}; and secondly, that these fcc interlayers influence the growth of the Co removing the hcp growth phase from the grains. The improved supercurrent in these samples may be due to the lack of stacking faults and dislocations compared to the other samples. An important open question from this work is why the Cu(5) interlayers reported by Khasawneh \textit{et al.} improve both $I_cR_N$ and the decay length of supercurrent, while the Pt(0.5) interlayers in this work improve only $I_cR_N$. Unfortunately the structural characterizations performed here do not offer any insight into this question.

To further support our interpretation of the data consider the asymmetric structures in FIG.~\ref{Cux}. It is reasonable to assume that a seed layer preceding Co growth is most important to improve the growth condition and atomic structure of the Co. FIG.~\ref{Cux} shows that indeed a higher $I_cR_N$ is only observed in the asymmetric sample with the bottom Pt(0.5) interlayer only. The Co $c$-plane spacing of these samples follows the same trend with $I_cR_N$, however is a less definitive result compared to the other samples. It is interesting to note that to recover the highest $I_cR_N$ observed in this work requires both interlayers, and that a decrease in $I_cR_N$ is observed for asymmetric samples with only the top Pt(0.5) interlayer. These observations in $I_cR_N$ and Co $c$-plane spacing suggest some importance of having uniform strain across the Josephson junction, but this hypothesis requires further study.





We next speculate why as the Pt layer is made thicker, the enhanced transmission of supercurrent decreases towards the value with no Pt interlayer (FIG.~\ref{Ptx}). We consider what happens to the number of Cooper pairs in the junction as the Pt interlayer is made thicker. This is described by the \textit{S--N} proximity effect. For a metal such as Cu with a very long electron mean-free-path, the lengthscale of this proximity effect, $\xi_{\text{Cu}}$, can be very long \cite{PhysRevB.63.064502}. Therefore in Cu interlayer samples there is little loss of Cooper pairs from the \textit{S--N} proximity effect. In Pt interlayer samples, the three important contributions to $\xi_{\text{Pt}}$ are the much shorter electron mean-free-path (compared to Cu), the role of the interfacial Rashba spin-orbit coupling, and the magnetic moment gained by Pt in proximity to Co \cite{PhysRevB.65.020405,rowan2017interfacial}. The influence of the first may be studied in Josephson junctions containing only Pt, however we expect the second and third to be more important in our experiment. The exact role of the spin-orbit coupling in our samples is unknown. A theoretical work on Josephson current through a diffusive normal wire with intrinsic Rashba spin-orbit coupling has been studied \cite{PhysRevB.93.024522}, however it is difficult to know how to apply these results to our experiment. The Pt magnetization will contribute to spin-flip scattering, especially if structural roughness propagates into a magnetic roughness, or there are competing interface anisotropies between Pt and Co. In any case, our results suggest that $\xi_{\text{Pt}}$ is short and hence any gain in supercurrent transmission from the addition of Pt interlayers will have to compete with the loss of Cooper pairs from this \textit{S--N} proximity effect.



Finally, we discuss our experiment in the context of theoretical predictions. Bergeret \textit{et al.} consider singlet-triplet conversion in the presence of Rashba and/or Dresselhaus spin-orbit coupling \cite{PhysRevB.89.134517}. They provide a criteria for generating the LRTC, namely that the vector operator $\big[ \mathcal{\hat{A}}_k,[ \mathcal{\hat{A}}_k, h^a \sigma^a ] \big]$ is \emph{not} parallel to the exchange field operator ($h^x\sigma^x, h^y\sigma^y, h^z\sigma^z$). They obtain the former in equation (67) of their work, which we simplify here for a metallic system (such as ours) with finite Rashba ($\alpha \neq 0$) and zero Dresselhaus ($\beta = 0$) contribution to the spin-orbit coupling as \cite{PhysRevB.89.134517} 
\begin{equation}
\big[ \mathcal{\hat{A}}_k,[ \mathcal{\hat{A}}_k, h^a \sigma^a ] \big] = 4 \alpha^2 (2 h^x \sigma^x + h^y \sigma^y + h^z \sigma^z ),
\end{equation}
where $\alpha$ is known in the literature as the Rashba constant and $\sigma$ is the vector of the Pauli matrices. In other words, (1) has components ($2h^x, h^y, h^z$). If the direction of the exchange field is perpendicular to the plane, ($h^x \neq 0$, $h^y = h^z = 0$), then there is no LRTC as (1) is parallel to the exchange field. Equally if $h^x = 0$ and $h^y = h^z \neq 0$ (in-plane magnetization), then there is no LRTC by the same logic. If $h^x$ and at least one of $h^y$ or $h^z$ are non-zero, then ($2h^x, h^y, h^z$) has a component perpendicular to the exchange field ($h^x, h^y, h^z$), hence the LRTC can be created. 

To address this limitation, Bergeret \textit{et al.} propose performing the experiment with a magnetization in-plane \textit{F} layer fabricated into a current in-plane (lateral) Josephson junction, which they show can recover LRTC generation \cite{PhysRevB.89.134517}. Lateral Josephson junctions containing half-metals are well established \cite{keizer2006spin, PhysRevB.82.100501,doi:10.1063/1.3681138}, however substituting transition metal \textit{F} layers into this geometry has proved experimentally difficult. Single crystal Co nanowires contacted by W electrodes display zero resistivity \cite{wang2010interplay}, and recently a Josephson current has been passed laterally through Co disks when the separation between \textit{S} electrodes is very small \cite{lahabi2017controlling}. Polycrystalline Co wires show promise, however a zero resistance state is not observed \cite{kompaniiets2014long,kompaniiets2014proximity}. Jacobsen \textit{et al.} suggest alternatively to use the current perpendicular-to-plane geometry (employed in this work) with a ferromagnetic alloy which has both in and out-of-plane magnetization components together with a source of spin-orbit coupling \cite{jacobsen2016controlling}. This is closer to our experiment, where a small out-of-plane remanent magnetization is observed in FIG~\ref{loops} (d). This out-of-plane anisotropy, however, is most likely limited to the Co/Pt interface as our samples have predominant in-plane anisotropy. This may explain why LRTC generation in our samples appears (at best) to be very poor. In future works we will replace the Co/Ru/Co SAF with a multilayer such as [Pd/Co]$_n$ \cite{PhysRevB.96.224515}, [Pt/Co]$_n$, or [Ni/Co]$_n$ \cite{PhysRevB.86.224506} where careful engineering of the layer thicknesses can promote the required canted magnetization \cite{shepley2015modification}.






\section{Conclusions} 

The major conclusions of this work may be summarized as follows. The growth of 0.5~nm Pt interlayers in Josephson junctions containing Co/Ru/Co ferromagnetic layers significantly enhances the transmission of supercurrent through the junction. The origin of this enhanced transmission is believed to be primarily from the fcc Pt being an effective seed layer for the growth of fcc Co, which we confirm with complimentary x-ray diffraction measurements. Although most of our junctions displayed a supercurrent decay length consistent with singlet superconductivity, for the thickest Co layers a small residual supercurrent is present which may have a longer decay length. This small residual supercurrent is the best evidence in this work for spin-obit coupling mediating singlet-triplet conversion.


The data associated with this paper are openly available from the University of Leeds data repository \cite{NoteY}.

\begin{acknowledgments}
We thank G. Burnell and F. S. Bergeret for many helpful discussions, B. Bi for help with fabrication using the Keck Microfabrication Facility, and R. Loloee, J. Glick and V. Aguilar for assistance with sample growth, fabrication, magnetometry and transport measurements particularly in the early stages of this project. This work was supported by the Marie Sk\l{}odowska-Curie Action ``SUPERSPIN'' (grant number: 743791).

\end{acknowledgments}

\bibliography{library}

\end{document}